\documentclass[letter,bibyear]{aa} 

\usepackage{epsfig}
\usepackage{amsmath}
\usepackage{subfigure}
\usepackage{natbib}
\usepackage{multirow}
\usepackage{color}
\usepackage{longtable}
\usepackage{graphicx}
\usepackage{epstopdf}
\usepackage{booktabs}
\usepackage{stackrel,amssymb,amsmath}
\usepackage{txfonts}

\newcommand{\jms}{J.~Mol.~Spectrosc.}

\begin{document}

\title{Discovery of the interstellar cyanoacetylene radical cation HC$_3$N$^+$\thanks{Based on
observations carried out with the Yebes 40m telescope (projects 19A003, 20A014, 20D023, 21A011, 21D005 and 23A024). The 40m
radio telescope at Yebes Observatory is operated by the Spanish Geographic Institute (IGN, Ministerio de Transportes y Movilidad Sostenible).}}

\author{
C.~Cabezas\inst{1},
M.~Ag\'undez\inst{1},
Y.~Endo\inst{2},
B.~Tercero\inst{3,4},
N.~Marcelino\inst{3,4},
P.~de~Vicente\inst{4}, and
J.~Cernicharo\inst{1}
}

\institute{Dept. de Astrof\'isica Molecular, Instituto de F\'isica Fundamental (IFF-CSIC),
C/ Serrano 121, 28006 Madrid, Spain \newline \email carlos.cabezas@csic.es, jose.cernicharo@csic.es
\and Department of Applied Chemistry, Science Building II, National Yang Ming Chiao Tung University, 1001 Ta-Hsueh Rd., Hsinchu 300098, Taiwan
\and Observatorio Astron\'omico Nacional (OAN, IGN), C/ Alfonso XII, 3, 28014, Madrid, Spain
\and Observatorio de Yebes (IGN), Cerro de la Palera s/n, 19141 Yebes, Guadalajara, Spain
}

\date{Received; accepted}

\abstract{We report the first identification in space of HC$_3$N$^+$, the simplest member of the family
of cyanopolyyne cations. Three rotational transitions with half-integer quantum numbers from $J$=7/2 to
11/2 have been observed with the Yebes 40m radio telescope and assigned to HC$_3$N$^+$, which has an inverted
$^2\Pi$ ground electronic state. The three rotational transitions exhibit several hyperfine components due
to the magnetic and nuclear quadrupole coupling effects of the H and N nuclei. We confidently assign the
characteristic rotational spectrum pattern to HC$_3$N$^+$ based on the good agreement between the
astronomical and theoretical spectroscopic parameters. We derived a column density of (6.0$\pm$0.6)$\times$10$^{10}$ cm$^{-2}$ and a rotational temperature of 4.5$\pm$1\,K. The abundance ratio between HC$_3$N and HC$_3$N$^+$ is 3200$\pm$320. As found for the larger members of the family of cyanopolyyne cations (HC$_5$N$^+$ and HC$_7$N$^+$), HC$_3$N$^+$ is mainly formed through the reactions of H$_2$ and the cation C$_3$N$^+$ and by the reactions of H$^+$ with HC$_3$N. In the same manner than other cyanopolyyne cations, HC$_3$N$^+$ is mostly destroyed through a reaction with H$_2$ and a dissociative recombination with electrons.}


\titlerunning{HC$_3$N$^+$ in TMC-1}
\authorrunning{Cabezas et al.}

\maketitle

\section{Introduction}

The cold dark cloud TMC-1 shows a rich and complex chemistry in which molecular ions play roles as important intermediates. These charged species can rapidly react with neutral partners or recombine with electrons to form other ionic and neutral molecules under astrophysical conditions \citep{Agundez2013}. TMC-1 is known to contain a variety of protonated species of abundant neutral molecules. In the past few years, the list has increased considerably with the discovery of the cations HC$_5$NH$^+$ \citep{Marcelino2020}, HC$_3$O$^+$ \citep{Cernicharo2020a}, HC$_3$S$^+$ \citep{Cernicharo2021a}, CH$_3$CO$^+$ \citep{Cernicharo2021b}, HCCNCH+ \citep{Agundez2022}, HCCS$^+$ \citep{Cabezas2022a}, HC$_7$NH$^+$ \citep{Cabezas2022b}, NC$_4$NH$^+$ \citep{Agundez2023}, C$_5$H$^+$ \citep{Cernicharo2022a}, H$_2$C$_3$H$^+$ \citep{Silva2023}, and HC$_5$N$^+$ and HC$_7$N$^+$ \citep{Cernicharo2024}. The interstellar identification of only a few of these protonated molecules has been performed through the direct comparison of the astronomical spectra with laboratory spectroscopic measurements. HC$_3$O$^+$ \citep{Cernicharo2020a}, HC$_3$S$^+$ \citep{Cernicharo2021a}, and CH$_3$CO$^+$ \citep{Cernicharo2021b} were observed in the laboratory thanks to their astronomical rotational frequencies, and only the cation H$_2$C$_3$H$^+$ \citep{Silva2023} was observed in the laboratory prior to its discovery in TMC-1.

These protonated species are considered 'short-lifetime' or transient molecules under the terrestrial physical conditions; hence, their spectroscopic characterisations require experiments with in situ generation, such as those employing electric discharges \citep{Cernicharo2020a,Cernicharo2021a}; glow discharges \citep{Cernicharo2021b}; or electron impact ionisation techniques \citep{Silva2023}. However, most of these positive ions have eluded characterisation in the laboratory, and therefore their identification in space has been carried out using quantum chemical calculations and direct spectroscopic analysis of high spectral resolution line surveys of molecular clouds, such as QUIJOTE\footnote{Q-band Ultrasensitive Inspection Journey to the Obscure TMC-1 Environment} \citep{Cernicharo2021c}. Examples of these kinds of identifications are the cations HC$_5$NH$^+$ \citep{Marcelino2020}, HCCNCH+ \citep{Agundez2022}, HCCS$^+$ \citep{Cabezas2022a}, HC$_7$NH$^+$ \citep{Cabezas2022b}, NC$_4$NH$^+$ \citep{Agundez2023}, C$_5$H$^+$ \citep{Cernicharo2022a}, and the recently detected cations HC$_5$N$^+$ and HC$_7$N$^+$ \citep{Cernicharo2024}. Hence, high-level quantum chemical calculations are the main driver for the identification of most of these species.

In this work, we present a new case of molecular spectroscopy done in space. Thanks to sensitive high spectral resolution observations of the cold dark cloud TMC-1 achieved with QUIJOTE, we have detected various groups of unidentified lines, revealing a complex hyperfine structure, which we assign to HC$_3$N$^+$. This species is the simplest member of the family of cyanopolyyne cations, for which two other species have also been observed in TMC-1, HC$_5$N$^+$ and HC$_7$N$^+$ \citep{Cernicharo2024}. As our observations of TMC-1 show narrow lines and high spectral resolution, it has been possible to resolve the hyperfine structure of the rotational lines of HC$_3$N$^+$ and hence to fully characterise the carrier and derive very precisely its spectroscopic parameters. Based on these constants and their agreement with those obtained by high-level quantum chemical calculations, we confidently assign the observed lines to HC$_3$N$^+$, and we report its first detection in space.

\section{Quantum chemical calculations}\label{calculations}

Several theoretical calculations for HC$_3$N$^+$ have been published
\citep{Zhang2012,Leach2014,Dai2015,Desrier2016,Gans2016,Steenbakkers2023}. However, we carried out quantum
chemical calculations in order to estimate the molecular parameters for HC$_3$N$^+$ not reported in the literature
before. The equilibrium molecular geometry, rotational constant, and the electric dipole moment were calculated using the spin-restricted coupled cluster method with single, double, and perturbative triple excitations and with an explicitly correlated approximation (RCCSD(T)-F12; \citealt{Knizia2009}) and all electrons (valence and core) correlated together
with the Dunning's correlation consistent basis sets with a polarised core-valence correlation triple-$\zeta$ for explicitly
correlated calculations (cc-pCVTZ; \citealt{Hill2010}). These calculations were carried out using the Molpro 2024.1
program \citep{Werner2024}. In addition, we calculated other parameters necessary to interpret the rotational spectrum. At the optimised geometry, we calculated the magnetic hyperfine constants for the non-zero nuclear spin nuclei, hydrogen and nitrogen in this case, which are the Fermi ($b_F$) and the dipole-dipole constants ($T_{aa}$ and $T_{bb}$). Additionally, the nuclear electric quadrupole constants for the nitrogen nucleus ($\chi_{aa}$ and $\chi_{bb}$) have also been calculated using the B3LYP hybrid density functional \citep{Becke1993} with the polarised valence triple-$\zeta$ basis set (cc-pVTZ; \citealt{Woon1993}). Harmonic frequencies were computed at the same level of theory to estimate the centrifugal distortion constant. These calculations were done using the Gaussian16 program \citep{Frisch2016}. Table~\ref{rotctes} summarises the results of these calculations.

The Frosch and Foley \citep{Frosch1952} hyperfine constants were derived from the dipole-dipole interaction tensor \textbf{T} and the Fermi contact constant b$_F$ -- both were obtained from quantum chemical calculation -- using the relations

\begin{eqnarray}
   b = {b_F - {c\over 3}} = {b_F - {1\over 2} T_{aa}},
\end{eqnarray}

\begin{eqnarray}
   c = {{3\over 2} T_{aa}},
\end{eqnarray}

\begin{eqnarray}
   d = { T_{bb} - T_{cc}}.
\end{eqnarray}

We utilised the approximate relation \citep{Townes1975}

\begin{eqnarray}
  c = 3(a-d)
\end{eqnarray}
\noindent
between $a$, $c$, and $d$ to estimate the $a$ constant as

\begin{eqnarray}
   a = {{c\over 3} + d} = {{3\over 2} T_{aa} + T_{bb}}.
\end{eqnarray}

The main contribution to the hyperfine coupling in the $^2\Pi_{3/2}$ state comes from the constant h$_1$, defined as

\begin{eqnarray}
h_1 = {a + {(b + c) \over 2}}.
\end{eqnarray}

\begin{table}
\small
\centering
\caption{Molecular constants (all in megahertz) of HC$_3$N$^+$ in $\tilde{X}$$^2\Pi_{3/2}$.}
\label{rotctes}
\centering
\begin{tabular}{{lcc}}
\hline
\hline
Constant       & TMC-1      & Theoretical$^a$  \\
\hline
$B$            &  4533.31711(148)$^b$  &      4535    \\
$D$            &   0.000456(36)        &    0.000469  \\
$b_F$(H)       &     -39.9(36)         &     -36.37   \\
$T_{aa}$(H)    &         -             &     22.20    \\
$T_{bb}$(H)    &         -             &     -3.87    \\
$a$(H)         &    18.53(160)         &     25.55    \\
$c$(H)         &      [33.3]$^c$       &      33.3    \\
$b_F$(N)       &     9.52(276)         &      0.13    \\
$T_{aa}$(N)    &        -              &     -20.55   \\
$T_{bb}$(N)    &        -              &     38.01    \\
$a$(N)         &    33.67(129)         &     45.18    \\
$c$(N)         &     [-30.82]$^c$      &     -30.82   \\
$eQq$(N)       &    -5.442(62)         &     -6.21    \\
|$\mu$| (D)    &        -              &        5.5   \\
\hline
\hline
\end{tabular}
\tablefoot{\\
\tablefoottext{a}{These values have been obtained from quantum chemical calculations at different levels of theory. (See text for details.)}
\tablefoottext{b}{The uncertainties (in parentheses) are in units of the last significant digits.}
\tablefoottext{c}{Values in brackets have been kept fixed to those obtained theoretically.}
}
\end{table}
\normalsize

\section{Astronomical observations}\label{astro_obs}

The observational data presented in this work are part of the QUIJOTE spectral line survey
\citep{Cernicharo2021c} in the Q-band towards TMC-1(CP) ($\alpha_{J2000}=4^{\rm h} 41^{\rm  m}
41.9^{\rm s}$ and $\delta_{J2000}=+25^\circ 41' 27.0''$) that was performed at the Yebes 40m radio
telescope. This survey was done using a receiver built within the Nanocosmos project\footnote{\texttt{https://nanocosmos.iff.csic.es/}} consisting of two cooled high electron mobility transistor (HEMT) amplifiers covering the 31.0-50.3 GHz band with horizontal and vertical polarisations. Fast Fourier transform spectrometers (FFTSs) with $8\times2.5$ GHz and a spectral resolution of 38.15 kHz provided the whole coverage of the Q-band in both polarisations. Receiver temperatures were 16\,K at 32 GHz and 30\,K at 50 GHz. The experimental setup has been described in detail by \citet{Tercero2021}.

All observations were performed using the frequency-switching observing mode with a frequency throw of 10 and 8 MHz. The total observing time on the source for data taken with frequency throws of 10 MHz and 8 MHz was 465 and 737 hours,
respectively. Hence, the total observing time on source was 1202 hours. The QUIJOTE sensitivity varies between 0.08 mK at 32 GHz and 0.2 mK at 49.5 GHz, and it is around 50 times better than that of previous line surveys in the Q-band
of TMC-1 \citep{Kaifu2004}. A detailed description of the line survey and the data analysis procedure followed are
provided in \citet{Cernicharo2021c,Cernicharo2022b}. The main beam efficiency can be given across the Q-band by $B_{\rm eff}$=0.797 exp[$-$($\nu$(GHz)/71.1)$^2$]. The forward telescope efficiency is 0.97. The telescope beam size at half-power intensity is 54.4$''$ at 32.4 GHz and 36.4$''$ at 48.4 GHz. The absolute calibration uncertainty is 10\,$\%$. The data were
analysed with the GILDAS package.\footnote{\texttt{http://www.iram.fr/IRAMFR/GILDAS}}

\begin{figure*}[]
\centering
\includegraphics[width=0.98\textwidth]{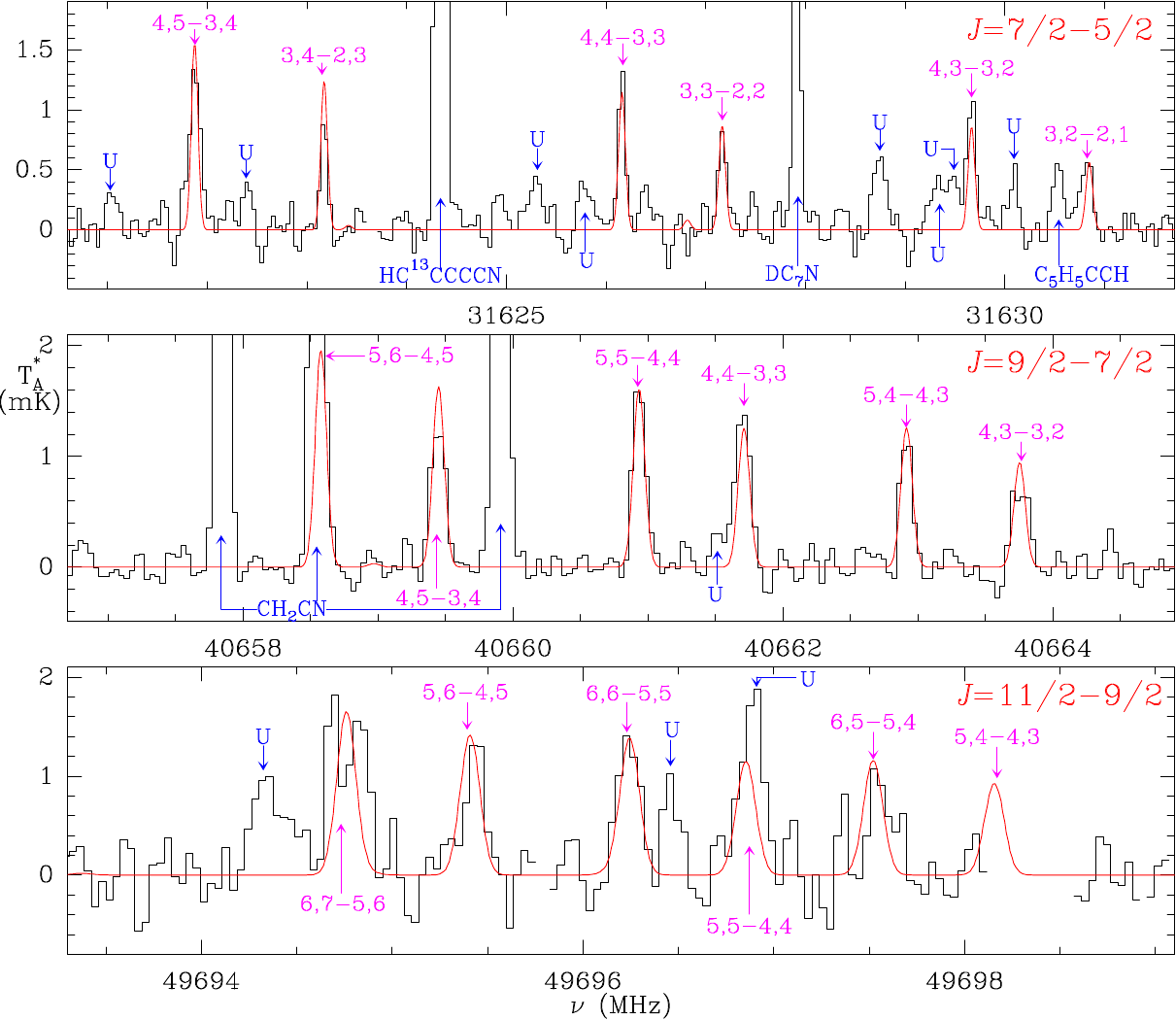}
\caption{Observed transitions of HC$_3$N$^+$ in TMC-1 with $F_1$ and $F$ quantum numbers indicated.
The abscissa corresponds to the rest frequency, and the ordinate is the antenna temperature corrected for atmospheric and telescope losses in milliKelvin. Blank channels correspond to negative features produced when folding the frequency-switched data. The red lines correspond to the synthetic spectra calculated for a column density of 6.0$\times$10$^{10}$ cm$^{-2}$ and
a rotational temperature of 4.5\,K.}
\label{fig_hc3n+}
\end{figure*}

\section{Results and discussion}\label{section_calculations}

\begin{table*}[h]
\centering
\caption{Observed line parameters for HC$_3$N$^+$.}
\label{line_parameters}
\begin{tabular}{ccccrcccc}
\hline
\hline
$J'$-$J$  & $F_1'$-$F_1$  & $F'$-$F$  &$\nu_{rest}$~$^a$ & (o-c)$^b$ & $\int T_A^* dv$~$^c$ &  $\Delta v$~$^d$ & $T_A^*$~$^e$& Notes\\
   &            &      &    (MHz)    & (MHz)                         &(mK\,km\,s$^{-1}$)    & (km\,s$^{-1}$)   &  (mK) &      \\
\hline
7/2-5/2  & 4-5 & 3-4 &   31621.877$\pm$0.010   &   0.002  &1.53$\pm$0.08& 1.07$\pm$0.07&1.33$\pm$0.10  \\
7/2-5/2  & 3-4 & 2-3 &   31623.173$\pm$0.010   &  -0.000  &1.05$\pm$0.07& 0.90$\pm$0.01&1.16$\pm$0.10  \\
7/2-5/2  & 4-4 & 3-3 &   31626.161$\pm$0.010   &   0.001  &1.07$\pm$0.13& 0.75$\pm$0.08&1.34$\pm$0.13&A\\
7/2-5/2  & 3-3 & 2-2 &   31627.170$\pm$0.010   &   0.005  &0.91$\pm$0.13& 0.99$\pm$0.18&0.87$\pm$0.10  \\
7/2-5/2  & 4-3 & 3-2 &   31629.665$\pm$0.010   &   0.000  &0.92$\pm$0.13& 0.92$\pm$0.18&0.94$\pm$0.10&B\\
7/2-5/2  & 3-2 & 2-1 &   31630.840$\pm$0.020   &  -0.006  &0.88$\pm$0.13& 0.95$\pm$0.23&0.61$\pm$0.10  \\
9/2-7/2  & 5-6 & 4-5 &                         &          &             &              &             &C\\
9/2-7/2  & 4-5 & 3-4 &   40659.446$\pm$0.010   &   0.002  &1.37$\pm$0.08&  0.97$\pm$0.07& 1.33$\pm$0.09 \\
9/2-7/2  & 5-5 & 4-4 &   40660.934$\pm$0.010   &   0.001  &1.44$\pm$0.08&  0.81$\pm$0.05& 1.67$\pm$0.09 \\
9/2-7/2  & 4-4 & 3-3 &   40661.701$\pm$0.010   &   0.009  &1.45$\pm$0.09&  0.98$\pm$0.07& 1.39$\pm$0.09 \\
9/2-7/2  & 5-4 & 4-3 &   40662.912$\pm$0.010   &  -0.004  &0.92$\pm$0.07&  0.72$\pm$0.06& 1.20$\pm$0.09 \\
9/2-7/2  & 4-3 & 3-2 &   40663.761$\pm$0.010   &  -0.008  &0.90$\pm$0.09&  1.10$\pm$0.09& 0.76$\pm$0.09 \\
11/2-9/2 & 6-7 & 5-6 &   49694.811$\pm$0.050   &  -0.049  &2.41$\pm$0.24&  1.51$\pm$0.18& 1.51$\pm$0.25  \\
11/2-9/2 & 5-6 & 4-5 &   49695.446$\pm$0.050   &   0.037  &1.00$\pm$0.19&  0.69$\pm$0.17& 1.38$\pm$0.25  \\
11/2-9/2 & 6-6 & 5-5 &   49696.225$\pm$0.050   &  -0.020  &1.21$\pm$0.19&  0.83$\pm$0.14& 1.43$\pm$0.25  \\
11/2-9/2 & 5-5 & 4-4 &   49696.847$\pm$0.050   &   0.008  &1.65$\pm$0.21&  0.85$\pm$0.13& 1.82$\pm$0.25&B  \\
11/2-9/2 & 6-5 & 5-4 &   49697.548$\pm$0.050   &   0.026  &0.98$\pm$0.22&  1.01$\pm$0.30& 0.91$\pm$0.25  \\
11/2-9/2 & 5-4 & 4-3 &                         &         &             &               &              &D\\
\hline
\hline
\end{tabular}
\tablefoot{
\tablefoottext{a}{Rest frequency assuming a v$_{LSR}$ of 5.83 km\,s$^{-1}$ \citep{Cernicharo2020c}.}
\tablefoottext{b}{Observed frequency minus calculated frequency from the fit.}
\tablefoottext{c}{Integrated intensity (in mK km\,s$^{-1}$).}
\tablefoottext{d}{Line width (in km\,s$^{-1}$).}
\tablefoottext{e}{Antenna temperature (in mK).}
\tablefoottext{A}{Data only from the set of data with frequency switching throw of 8 MHz.}
\tablefoottext{B}{Partially blended with a U feature.}
\tablefoottext{C}{The line is fully blended with a strong line of CH$_2$CN.}
\tablefoottext{D}{Affected by a negative feature produced in the folding of the frequency switching data.}
}
\end{table*}

Since HC$_5$N$^+$ and HC$_7$N$^+$ have been detected in TMC-1, the smaller member of this family of cyanopolyyne cations is expected to be present as well in this molecular cloud. A first attempt was made to find the rotational signatures of HC$_3$N$^+$ in our data, based on the estimation of the $B$ rotational constant reported in several theoretical works \citep{Cernicharo2024}. However, we could not find unidentified lines with frequencies in harmonic relation attributable to HC$_3$N$^+$ in the explored frequency ranges. Hence, we concluded that the abundance of HC$_3$N$^+$ in TMC-1 should be $\le$10$^{11}$ cm$^{-2}$.

Similarly to HC$_5$N$^+$ and HC$_7$N$^+$, HC$_3$N$^+$ has an inverted $^2\Pi$ electronic ground state with the
$\Omega$=3/2 ladder lower in energy than the $\Omega$=1/2 ladder \citep{Gans2016}. Consequently, the three rotational transitions $J$=7/2-5/2, $J$=9/2-7/2, and $J$=11/2-9/2 that are within the frequency coverage of QUIJOTE will exhibit significant hyperfine structure due to electric and magnetic effects caused by the interaction between H and N nuclear spin and electron orbital and electron spin angular momenta. As it occurs for HC$_5$N$^+$, the $\Lambda$-doubling components are not expected to be resolved at the spectral resolution of our observations due to the values of $\Delta$($^2\Sigma - ^2\Pi$) -- the energy difference between the $^2\Sigma$ and the $^2\Pi$ electronic states -- and $A_{SO}$ (the spin-orbit constant), which are 1.914 eV \citep{Leach2014} and $-$44$\pm$2\,cm$^{-1}$ \citep{Gans2016}, respectively. We predicted the rotational spectrum of HC$_3$N$^+$ in the 30-50 GHz region using the theoretical values for the molecular constants shown in Table \ref{rotctes}. The predicted pattern for the most intense lines consists of six lines with a separation of 1.2-1.6 MHz.
This can be seen as a triplet of doublets produced by the nitrogen and hydrogen nuclei, respectively. We found a group of unidentified lines at 31.6 GHz, which is only 15 MHz from the predicted frequencies of the $J$=7/2-5/2 transition and shows a comparable hyperfine structure. The lines, with intensities ranging from 0.5 mK to $\sim$1.2 mK, are mixed with stronger transitions of other species such as HC$^{13}$CCCCN or DC$_7$N, as shown in Fig.\ref{fig_hc3n+}. The frequency separation between the six unidentified lines is larger than predicted, which indicates that the values of the $b_F$ constants of H and N nuclei must be larger too. Another group of unidentified lines was found in the 40.6 GHz frequency region, 30 megahertz from the frequency predicted for the $J$=9/2-7/2 transition, showing a similar pattern to the lines at 31.6 GHz. In this case, the group of lines consists of five lines with intensities between 0.5 mK and $\sim$1.6 mK that appear close to the transitions of the CH$_2$CN radical. In fact, it seems that one of the HC$_3$N$^+$ hyperfine components could be fully
blended with a strong line from CH$_2$CN. Assuming that these two groups of lines correspond to HC$_3$N$^+$, we  accurately predicted the frequency for the $J$=11/2-9/2 transition, and we also found several weak lines at 49.7 GHz. This frequency region is the least sensitive of our QUIJOTE data, and those lines are strongly affected by the noise. Nonetheless, the expected hyperfine pattern can be easily distinguished, as seen in Fig.\ref{fig_hc3n+}. The derived frequencies, intensities, and line widths of the observed lines are given in Table~\ref{line_parameters}.

The observed frequencies were analysed with the SPFIT program \citep{Pickett1991} using an effective Hamiltonian
for a linear molecule in a $^2\Pi$ electronic state. The employed Hamiltonian has the following form:
\begin{equation}
\textbf{H}_{eff} = \textbf{H}_{rot} + \textbf{H}_{mhf} + \textbf{H}_Q,
\end{equation}
where \textbf{H}$_{rot}$ contains rotational and centrifugal distortion parameters,
$\textbf{H}_{mhf}$ represents the magnetic hyperfine coupling interaction term due to the N and
H nuclei, and $\textbf{H}_Q$ represents the nuclear electric quadrupole interaction due to the N nucleus.
The coupling scheme used was \textbf{F}$_1$\,=\,\textbf{J}\,+\,\textbf{I}$_1$ and
\textbf{F}\,=\,\textbf{F}$_1$\,+\,\textbf{I}$_2$, where \textbf{I}$_1$\,=\,\textbf{I}(H)
and \textbf{I}$_2$\,=\,\textbf{I}(N).

We considered uncertainties of 10 kHz for the $J$=7/2-5/2 and $J$=9/2-7/2 rotational transitions, while
the uncertainty assigned to the components of the $J$=11/2-9/2 transition was 50 kHz, due to the lower signal-to-noise ratio of the lines in the 49.6 GHz region. The results obtained from the fit are shown in Table~\ref{rotctes}. We experimentally determined the values of the rotational constant, $B$; the centrifugal distortion constant, $D$; the magnetic hyperfine constants due to the nitrogen and hydrogen nuclei, $a$ and $b_F$; and the nuclear electric quadrupole constant of the nitrogen nucleus, $eQq$. The $c$ magnetic hyperfine constants of nitrogen and hydrogen nuclei were kept fixed to the theoretical values calculated in this work. The standard deviation of the fit is 18.2\,kHz, about a half of the frequency resolution in our observations.

The identification of HC$_3$N$^+$ as the carrier of the observed lines is unequivocal. The experimental rotational constant $B$ deviates from the theoretical value by just 0.03\,\%. The level of theory employed in this work, RCCSD(T)-F12/cc-pCVTZ-F12, also reports very accurate values for the rotational constants of the neutral counterpart HC$_3$N molecule. Although the prediction of the centrifugal distortion constant is usually less accurate, the derived value from our fit also agrees very well with that predicted by quantum chemical calculations. These two constants provide information on the mass distribution of the molecular species, and thus, the good accordance between the observed and calculated values could serve by itself to identify the spectral carrier. However, the high resolution of our survey allowed us to scrutinise the fine and/or hyperfine structure of the rotational transitions. This structure depends strongly on the electronic environment of the molecular species, which is related to the number of nuclei with a non-zero nuclear spin and their relative positions. Hence, the fine and hyperfine coupling constants determined are ideal fingerprints of the spectral carrier, providing key information for its identification. As can be seen in Fig.\ref{fig_hc3n+}, the spectral pattern for the three transitions is very well described by the employed Hamiltonian, not only in terms of frequencies but also in relation to the relative intensities of each hyperfine component. Regarding this point, it should be noted that the hyperfine components may be affected by both negative components produced when folding the frequency-switched data and emissions of some other still unidentified chemical species. The limited number of rotational transitions observed and the fact that only one spin component, $\Omega$=3/2, is observed makes it so that only the values for two hyperfine constants for each nuclei can be determined, and not very accurately. Nevertheless, the values of $a$ and $b_F$ for nitrogen and hydrogen nuclei are in agreement with those predicted theoretically. In addition, the $eQq$ constant for the nitrogen nucleus is also in accordance with that obtained by quantum chemical calculations. The $eQq$ constant is a sensitive measure of the electric field gradient in a molecule. The experimental value, $-$5.394 MHz, is larger than that found for the neutral counterpart HC$_3$N ($eQq$=$-$4.31 MHz), and this is compatible with a cationic species with a terminal nitrogen. This fact has been observed before for NNH$^+$ \citep{Caselli1995} and NC$_2$NH$^+$ \citep{Gottlieb2000}. In light of all of these arguments, we definitively conclude that the carrier of the unidentified lines observed in TMC-1 is HC$_3$N$^+$.

Using the derived molecular constants from Table \ref{rotctes} and adopting for HC$_3$N$^+$ the dipole moment
of 5.5\,D derived in this work (see Table \ref{rotctes}), we derived a rotational temperature of 4.5$\pm$0.5\,K and a column density of (6.0$\pm$0.6)$\times$10$^{10}$ cm$^{-2}$. We assumed a source of uniform brightness with a diameter of 80$''$. The derived rotational temperature is not well constrained, as the energies of the upper levels of the three observed transitions range between 4 and 9\,K. The derived column density is almost two times smaller than the upper limit estimated by \citet{Cernicharo2024}. The derived rotational constant of HC$_3$N$^+$, $B\sim$4533 MHz, is 15 MHz below the lower range explored for $B$ by \citet{Cernicharo2024}.

Adopting the column density derived for HC$_3$N of 1.9$\times$10$^{14}$ cm$^{-2}$ \citep{Tercero2024},
the HC$_3$N/HC$_3$N$^+$ abundance ratio is 3200$\pm$320, which is a factor of five larger than the ratio between
HC$_5$N and HC$_5$N$^+$ and a factor of three larger than that of HC$_7$N and HC$_7$N$^+$
\citep{Cernicharo2024}. Adopting the column density derived for C$_3$N of 1.8$\times$10$^{13}$ cm$^{-2}$ \citep{Cernicharo2020b}, the C$_3$N/HC$_3$N$^+$ abundance ratio is 300$\pm$30, while the C$_5$N/HC$_5$N$^+$ abundance ratio was found to be 4.8$\pm$0.8 \citep{Cernicharo2024}. These results indicate that there are significant differences in the formation and destruction paths of these cations.
The cation HC$_3$N$^+$ can be seen as the protonated form of C$_3$N, whose proton affinity has been calculated to be 772  kJ mol$^{-1}$. Even though this value is a bit smaller than those calculated for C$_5$N and C$_7$N, the reactions of the proton transfer from abundant cations such as HCO$^+$ are exothermic and can provide a formation pathway to HC$_3$N$^+$ in TMC-1. Nevertheless, according to our previous chemical modelling calculations \citep{Cernicharo2024}, HC$_3$N$^+$ is mostly formed by the reactions H$_2$ + C$_3$N and H$^+$ + HC$_3$N. The abundance calculated in that work for HC$_3$N$^+$ is low, about two orders of magnitude lower than the value derived here, due to the non-negligible reactivity with H$_2$ \citep{Knight1985}. Therefore, it would be interesting to investigate the low-temperature chemical kinetics of HC$_3$N$^+$, HC$_5$N$^+$, and HC$_7$N$^+$ with H$_2$ in order to better understand the formation and survival of these highly unstable radical cations in TMC-1. The cation HCN$^+$ ($^2\Pi_i$) is worth being searched for, but its first transition, $J$=5/2-3/2, is outside the frequency ranges of our surveys.

\section{Conclusions}

We report the discovery of the radical cation HC$_3$N$^+$ in TMC-1. Thanks to the sensitive high spectral resolution observations of the QUIJOTE survey, we have observed three rotational transitions for this molecule. The transitions reveal a complex hyperfine structure, which was fully analysed and assigned to HC$_3$N$^+$, allowing us to derive its rotational spectroscopic parameters for the first time. This represents an invaluable alternative to overcome the lack of laboratory experimental data for very unstable molecules. The column density derived for HC$_3$N$^+$ is (6.0$\pm$0.6)$\times$10$^{10}$ cm$^{-2}$, and the rotational temperature is 4.5$\pm$1\,K. The abundance ratio between HC$_3$N and HC$_3$N$^+$ is 3200$\pm$320.

\begin{acknowledgements}

We thank Ministerio de Ciencia e Innovaci\'on of Spain (MICIU) for funding support through projects PID2019-106110GB-I00, and PID2019-106235GB-I00. We also thank ERC for funding through grant ERC-2013-Syg-610256-NANOCOSMOS and the Consejo Superior de Investigaciones Cient\'ificas (CSIC; Spain) for funding through project PIE 202250I097. C.C., M.A, and J.C. thank Ministerio de Ciencia e Innovaci\'on of Spain (MICIU) for funding support through projects PID2019-106110GB-I00, PID2019-107115GB-C21 / AEI / 10.13039/501100011033, and PID2019-106235GB-I00. The present study was also supported by Ministry of Science and Technology of Taiwan and Consejo Superior de Investigaciones Cient\'ificas under the MoST-CSIC Mobility Action 2021 (Grants 11-2927-I-A49-502 and OSTW200006).

\end{acknowledgements}

\normalsize
\begin{appendix}

\end{appendix}
\end{document}